\documentstyle[12pt]{article}
\begin{document}
\title{NO SUPERLUMINAL EXPANSION OF THE UNIVERSE\thanks{Alberta
Thy-13-93,
gr-qc/9303008}}
\author{ Don N. Page\\CIAR Cosmology Program\\Theoretical Physics
Institute\\
Department of Physics\\University of Alberta\\Edmonton,
Alberta\\Canada T6G
2J1\\
Internet:  don@page.phys.ualberta.ca}
\date{1993 March 2}
\maketitle
\large
\begin{abstract}
\baselineskip 20pt
The apparent superluminal growth in the size of a sufficiently large
part of
the universe can be ascribed to the special relativistic effect of
time
dilation.
\end{abstract}
\normalsize
\pagebreak
\baselineskip 24.2pt

In flat (Minkowski) spacetime, special relativity forbids signals and
material
objects from moving faster than the speed of light $c$.  In any
inertial
(Lorentz) frame, the linear size (length) of an object cannot expand
faster
than $2c$, the limit being given by the case in which the two ends of
the
object each move directly apart at the speed of light.  Any faster
growth would
be superluminal and would not be allowed by special relativity for an
object
with definite material ends.

(If the material defining the ends can change, there is no such
limitation.
For example, if a lake freezes over nearly simultaneously, the size
of the ice
could grow arbitrarily rapidly.  As a function of time while the
water freezes,
the ends of the ice are not definite water molecules but rather
growing
sequences of molecules  which at each instant are at the two ends of
the ice.
Another example could be the size of a shadow.  We shall henceforth
exclude
such cases from consideration and only consider objects with definite
ends that
cannot move faster than light as seen locally.)

On the other hand, a sufficiently large region of the universe, with
ends that
are definite physical entities such as elementary particles or
galaxies,
apparently can expand much faster than $2c$.  For example, a
spatially flat
homogeneous isotropic cosmological model has the $k=0$
Friedmann-Robertson-Walker metric
	\begin{equation}
	ds^2 = -c^2 dt^2 + a^2 (t) (dx^2 + dy^2 + dz^2).
	\end{equation}
The particles in this idealized model (which remarkably seems to be
an
excellent approximation for our universe on length scales of one to
ten or so
billion light years) each stay at fixed comoving coordinates
$(x,y,z)$ as their
proper time $t$ increases.  Thus the particles are comoving with the
spatial
coordinates and hence may be considered to be locally at rest with
respect to
these coordinates.

However, the physical distance between two particles, as measured
along a
geodesic of a hypersurface of constant $t$, is
	\begin{equation}
	D_t = D_c a(t)
	\end{equation}
where
	\begin{equation}
	D_c = \sqrt{\Delta x^2 + \Delta y^2 + \Delta z^2}
	\end{equation}
is the (constant) comoving coordinate distance between the two
comoving
particles, with $(\Delta x, \Delta y, \Delta z)$ being the
differences beween
their comoving coordinates.  The physical distance grows at the rate
	\begin{equation}
	v_t = \frac{d}{dt}D_t = D_c \frac{da}{dt} = H D_t,
	\end{equation}
where
	\begin{equation}
	H = \frac{1}{a} \frac {da}{dt}
	\end{equation}
is the Hubble expansion rate of the universe at the time t, the
logarithmic
rate of linear expansion of any comoving part (a part with fixed
comoving
coordinates $x,y,z$), with the dimension of reciprocal time.

For the spatially flat model (1) (as also for any open cosmological
model),
there is no upper limit to the coordinate distance $D_c$, and hence
also to the
physical distance $D_t$, between particles at any fixed $t$.  Thus
there is no
upper limit to the speed $v_t$ of expansion given by Eq. (4).  In
particular,
it can be much larger than the usual special relativistic limit $2c$.
How is
this possible?

One idea is that the particles are not actually moving apart, but
that space
between the particles is expanding by a general relativistic (curved
spacetime)
effect.  This interpretation is certainly natural for the
cosmological model
(1) written in terms of comoving coordinates, but is it distinct from
ordinary
expansion in special relativity (flat spacetime)?

To test this, we can use comoving coordinates to examine the
expansion of the
distance between particles in the flat Minkowski spacetime of special
relativity.  In inertial coordinates $(T,X,Y,Z)$, the metric of flat
spacetime
is
	\begin{equation}
	ds^2 = -c^2 dT^2 + dX^2 + dY^2 + dZ^2.
	\end{equation}
Suppose we consider a set of particles which fill the region $|X|
\leq |T|$ and
which have fixed
	\begin{eqnarray}
	x & = & \tanh^{-1}\frac{X}{cT} = \frac{1}{2} \ln
\frac{cT+X}{cT-X}, \\
	y & = & Y, \\
	z & = & Z,
	\end{eqnarray}
which serve as comoving coordinates for these particles.  At $T=0$,
the
particles are all confined to the $(Y,Z)$ plane $(X=0)$, but then
each spreads
out perpendicularly from there with constant velocity
	\begin{equation}
	v = \frac {X}{T} = c \tanh x.
	\end{equation}
The proper time along each particle worldline is
	\begin{equation}
	t = \sqrt{1-v^2/c^2} T = \sqrt{T^2-X^2/c^2}.
	\end{equation}

If we invert the relations (7)-(9) and (11), we get the inertial
coordinates
\linebreak[4] $(T,X,Y,Z)$ in terms of the comoving coordinates
$(t,x,y,z)$:
	\begin{eqnarray}
	T & = & t \cosh x, \\
	X & = & ct \sinh x, \\
	Y & = & y, \\
	Z & = & z.
	\end{eqnarray}
When we differentiate these and insert them into the Minkowski metric
(6), we
get the metric
	\begin{equation}
	ds^2 = -c^2 dt^2 + c^2 t^2 dx^2 + dy^2 + dz^2
	\end{equation}
for the $|X| \leq |T|$ region of the same flat spacetime, but written
in the
comoving coordinates of the spreading particles.

In these comoving coordinates, the physical distance between two
particles with
the same $y$ and $z$, as measured along a geodesic (constant $y$ and
$z$) of a
hypersurface of constant $t$, is (assuming $\Delta x \geq 0$)
	\begin{equation}
	D_t = c t \Delta x.
	\end{equation}
This grows at the rate
	\begin{equation}
	v_t = \frac{dD_t}{dt} = c \Delta x.
	\end{equation}
Again, this has no upper limit (e.g., $2c$), since $\Delta x$ can be
arbitrarily large.  We seem to have the possibility of an arbitrarily
rapid
expansion even within the special relativity of flat spacetime!  What
is going
wrong?

The first clue to what is happening is the fact that the hypersurface
of
constant $t$ (constant proper time from the two-surface $T=0, X=0$)
is not an
(extrinsically) flat hypersurface in the flat spacetime.  That is,
Eq. (11)
shows that the $t=const$ surface is curved in terms of the inertial
coordinates
$(T,X,Y,Z)$.  Thus geodesics of that hypersurface are not necessarily
geodesics
of the enveloping spacetime; the hypersurface $t=const.$ forces any
geodesic
within it to bend in the $(T,X)$ plane and therefore not be a
geodesic of the
full spacetime (unless the geodesic has $x$, and hence both $T$ and
$X$,
constant).  This implies that the physical distance $D_t$ within the
constant
$t$ hypersurface, given by Eq. (17), is not the geodesic distance
within the
spacetime.

  This situation is somewhat analogous to measuring distance on the
earth along
a great circle, which is a geodesic of the surface (idealized as a
sphere), but
not of three-dimensional space.  In particular, the distance on the
surface
between the poles is $\pi R$, whereas on a geodesic through the space
in the
middle it is $2R$, where $R$ is the radius of the earth.  If the
earth were
expanding at the speed of light, $dR/dt = c$, the geodesic distance
through the
center would be growing at the special relativistic limit $2c$, but
the great
circle distance on the surface would grow at $\pi c$, exceeding the
limit.
Thus we can easily get superluminal expansions if we measure
distances that are
not geodesics of the whole spacetime.

 However, if we go back to our special relativistic model above, we
find that
the spacetime geodesic distance between the two events at which the
two
particles (with the same $y$ and $z$) have the same proper time $t$
is
 	\begin{equation}
	D = \sqrt{\Delta X^2 - c^2 \Delta T^2} = 2 c t \sinh
\frac{1}{2} \Delta x,
	\end{equation}
which is {\it greater} than the curved distance $D_t$ given by Eq.
(17),
increasing the superluminal expansion.  How is it possible that the
curved
path, within the bent $t=const.$ hypersurface, could be shorter?

In a Riemannian space, with a positive-definite metric (so that all
directions
are spacelike rather than timelike), the shortest curve between two
points is a
geodesic, so it cannot be shortened.  But in a pseudo-Riemannian
space, such as
spacetime with its indefinite metric with timelike as well as
spacelike
directions, any spacelike curve can be shortened by bending it in the
timelike
direction so that it is more nearly lightlike or null (in the limit
of which it
would have zero length).  The geodesics of the $t=const.$
hypersurface, which
have length $D_t$, are curved in the timelike direction and hence are
shorter
than the geodesics of the enveloping spacetime, which have length
$D$.  The
same would be true in the Friedmann-Robertson-Walker metric (1) as in
the flat
metric (16).

Thus our first clue led to a correction (using the length of
geodesics of
spacetime rather than of a particular spatial hypersurface) that
worsens the
problem of superluminal expansion.  But now that we have a preferred
curve (the
spacetime geodesic) joining two points on the two particle
worldlines, we can
ask directly how fast these worldlines are moving apart.  If the
worldlines are
orthogonal to the geodesic joining them, another geodesic joining two
slightly
later points on the worldlines will have the same length (to first
order in the
time separation of the two successive joining geodesics), so then the
geodesic
length between the worldlines is not expanding.  This is as true in
the curved
spacetime of general relativity as in the flat spacetime of special
relativity,
since it is essentially a consequence of the fact that a geodesic is
a curve of
extremal length (meaning that with fixed endpoints, its length does
not change
to first order in a perturbation of the curve, though the extremum
need not be
a maximum or a minimum and indeed is not for a spacelike geodesic,
which gives
a saddle point for the distance).  Therefore, whether in flat
spacetime or in
curved spacetime, in no case does the space expand between two
worldlines which
are orthogonal to geodesics joining them.

If the worldlines are not orthogonal to a succession of geodesics
joining
corresponding points along them, then the rate of expansion is given
by the sum
of the outward components of the two worldline velocities (the
projections
parallel to the geodesic joining the worldlines at the time the
expansion is
measured), whether in flat spacetime or in curved.  Here we finally
come to the
crucial question, which is whether these velocities and rates are
with respect
to time $T$ measured by observers moving orthogonally to the joining
geodesic,
or with respect to the proper time $t$ of the worldlines themselves.
It is the
former observers that would see the geodesic as existing in one
simultaneous
moment and whose geodesic length is a purely spacelike distance.  The
time $T$
of the observers orthogonal to the joining geodesic will be dilated
with
respect to the proper time $t$ of the particle worldlines by the
relativistic
``gamma" factor
	\begin{equation}
	\gamma = \frac{dT}{dt} = (1-\frac{v^2}{c^2})^{-1/2} \geq 1,
	\end{equation}
where $v$ is the relative speed of the particle worldlines with
respect to the
orthogonal observers.

It is the derivative of the spacetime geodesic length $D$ with
respect to the
time $T$ of observers moving orthogonally to that geodesic at its
ends that is
the sum of the outward components of the ordinary 3-velocities of the
ends with
respect to the orthogonal abservers and hence cannot be larger than
twice the
speed of light, the luminal limit
	\begin{equation}
	\frac{dD}{dT} = {\bf v_1 \cdot n_1 + v_2 \cdot n_2} < 2c,
	\end{equation}
where ${\bf v_1}$ and ${\bf v_2}$ are the ordinary 3-velocities of
the ends,
and ${\bf n_1}$ and ${\bf n_2}$ are the unit outward spacelike normal
vectors
along the joining geodesic.  This inequality applies to curved
spacetime at
well as to flat spacetime.  Therefore, if $dD/dT$ is used as the
definition of
the expansion rate of any region of the universe that has definite
physical
entities at the ends (so that the ends cannot move faster than light
locally),
that region, no matter how large, cannot expand superluminally.

On the other hand, the derivative of $D$ with respect to the proper
time $t$ of
the particle worldlines at the ends of the joining geodesic is the
sum of the
outward components of the 4-velocities ${\bf u_1}$ and ${\bf u_2}$ of
the ends,
	\begin{equation}
	\frac{dD}{dt} = {\bf u_1 \cdot n_1 + u_2 \cdot n_2}
	= \gamma_1 {\bf v_1 \cdot n_1} + \gamma_2 {\bf v_2 \cdot
n_2},
	\end{equation}
which can exceed $2c$.  Similarly, even though the physical distance
$D_t$
along a geodesic of the (curved) spacelike hypersurface of fixed $t$
is
generally smaller than $D$, its derivative with respect to the proper
time $t$
can exceed $2c$, in which case one has
	\begin{equation}
	2c < \frac{dD_t}{dt} < \frac{dD}{dt} < (\gamma_1 + \gamma_2)
c.
	\end{equation}

Thus we see that $(\gamma_1 + \gamma_2)c$, rather than $2c$, is the
luminal
limit when the expansion rate is measured with respect to the proper
time $t$
of the ends, as one normally does in calculating cosmological
expansion rates,
such as $v_t$ given by Eq. (4) above.

In conclusion, we see that no region of the universe, with definite
physical
entities as ends, can expand faster than $2c$ with respect to the
time as
measured by observers moving orthogonally to the spacetime geodesics
joining
the two ends.  With respect to the proper time of the ends, the
expansion can
in principle be as large as $(\gamma_1 + \gamma_2)c$, which is larger
than
$2c$, but this is simply the time dilation effect of special
relativity, just
as applicable in the curved spacetime of the universe as in flat
spacetime.

This work was  supported in part by the Natural Sciences and
Engineering
Research Council of Canada.

\end{document}